\documentstyle[psfig,conf_iap,10pt]{article}
\begin{document}
\newcommand{\lya}{Ly$\alpha$ }
\newbox\grsign \setbox\grsign=\hbox{$>$} \newdimen\grdimen \grdimen=\ht\grsign
\newbox\simlessbox \newbox\simgreatbox
\setbox\simgreatbox=\hbox{\raise.5ex\hbox{$>$}\llap
     {\lower.5ex\hbox{$\sim$}}}\ht1=\grdimen\dp1=0pt
\setbox\simlessbox=\hbox{\raise.5ex\hbox{$<$}\llap
     {\lower.5ex\hbox{$\sim$}}}\ht2=\grdimen\dp2=0pt
\newcommand{\simgt}{\mathrel{\copy\simgreatbox}}
\newcommand{\simlt}{\mathrel{\copy\simlessbox}}

\heading{%
%
Metal Lines in Cosmological Models of \lya Absorbers 
%
} 
\par\medskip\noindent
\author{%
Uffe Hellsten$^1$, Lars Hernquist$^1$, Neal Katz$^2$, David H. Weinberg$^3$
}
\address{%
University of California, Lick Observatory, Santa Cruz, CA 95064
}
\address{%
University of Mass., Dept. of Physics and Astronomy, Amherst, MA, 01003
}
\address{%
Ohio State University, Dept. of Astronomy, Columbus, OH 43210
}

\begin{abstract}
The metal absorption lines found in association with \lya absorbers of moderate to 
high HI column density contain valuable information about the metallicity and 
ionization conditions within the absorbers and offer a stronger test of models
of the intergalactic medium at $z\sim 3$ than HI absorption lines alone.

We have developed a method to predict the strengths of metal absorption
lines within the framework of cosmological models for the \lya forest.
The method consists of evaluating a quantity, the {\em Line Observability
Index}, for a database of hundreds of candidate metal lines, allowing a
comprehensive identification of the lines the model predicts to be
detectable associated with a \lya absorber of a given {\rm HI} column
density and metallicity.

Applying this technique to a particular class of models at $z\sim 2-4$, we predict 
that the OVI(1032 \AA, 1038 \AA) doublet is the only practical probe of the metallicity of 
low column density absorbers ($N_{\rm HI} \simlt 10^{14.5} \, {\rm cm}^{-2}$),
that CIV (1548 \AA) is the strongest line with rest wavelength
$\lambda_r > 1216$ \AA{} regardless of $N_{\rm HI}$, and that the strongest
metal lines should be CIII(977 \AA) and SiIII(1206.5 \AA), which peak at
$N_{\rm HI} \sim 10^{17} \, {\rm cm}^{-2}$.

\end{abstract}
\section{Introduction}
The prevailing interpretation of the \lya forest is that it arises naturally from
absorption by trace amounts of neutral hydrogen in a photo-ionized, inhomogenous 
intergalactic medium (IGM). To model the \lya forest within this picture no speculative
assumptions about the physical properties of individual \lya absorbers have to
be made. One merely considers a cosmological simulation of structure formation, including 
dark matter and a baryonic component as well as a photo-ionizing background radiation 
field, and evaluates the absorption properties along lines of sight through the simulation
box at desired redshifts. During the past few years it has been realized that the 
artificial absorption spectra resulting from this approach bear close
resemblance to the \lya forest seen in QSO absorption spectra, and these cosmological
models of the \lya forest have been able to quantitatively account for the observed
distribution functions in HI column density and b-parameters to a reasonable accuracy
(\cite{cen94} \cite{zha95} \cite{pet95} \cite{her96}; for related semianalytic modeling 
see, e.g., \cite{bi93} \cite{bi97} \cite{hui97}). 

Within recent years, observations have demonstrated that strong \lya forest absorbers
generally show associated metal line absorption (\cite{mey87} \cite{cow95} \cite{wom95}
\cite{son96}). Selected metal absorption lines can be readily incorporated into cosmological
models of the \lya forest from a knowledge of densities, temperature, and UV radiation 
field along the lines of sight, if a metal enrichment pattern of the baryonic IGM is
specified. Such models account fairly well for the observed properties of the CIV(1548 \AA,
1550 \AA) doublet and a handful of other lines, for an IGM metallicity $Z \sim 10^{-2.5} Z_{\odot}$,
if a scatter of about an order of magnitude is assumed (\cite{hae96} \cite{rau96} \cite{hel97a}).

Instead of making an a priori selection of a few metal lines to include in a model, it
is useful to make the models {\em predict} which metal lines, out of
hundreds of candidates, should be observable (in spectra of a given resolution and S/N)
associated with \lya absorbers of given metallicities and HI column densities.
Such an approach allows a comprehensive screening for lines that deserve a more detailed
treatment, dependent on the specific purpose of the modeling, and allows for the sharpest
possible test of the models. We describe the results from an implementation of such a technique 
in the following. 
\section{Line Observability Index for 199 candidate metal lines}
Let us denote a metal absorption line produced by an element $Z$ in ionization stage
$i$ with rest transition wavelength $\lambda$ and oscillator strength $f$ as 
$Z_{\lambda,f}^i$.
For such a line associated with with a \lya absorber of neutral hydrogen
column density $N_{\rm HI}$ and metallicity $[Z/H] \equiv \log{(n_Z/n_H)}-\log{(n_Z/n_H)_{\odot}}$
we can define the following {\em line observability index} :

\begin{eqnarray}
{\rm LOX}(Z_{\lambda , f}^i,[{\rm Z/H}],N_{\rm HI}) & \equiv &  -17.05 \, + \,
\log{N_{\rm HI}} \, + \, [{\rm Z/H}] \, + \log{(f\, {\lambda}^2)} \\ \nonumber 
 & & +\log{({n_Z/n_{\rm H}})_{\odot}} \, + \,
\log{(x_Z^i/x_H^0)}
\end{eqnarray}
(see \cite{hel97b} for more details). This expression assumes units of ${\rm cm}^{-2}$ for
$N_{\rm HI}$ and \AA{} for $\lambda$. The choice of additive constant then implies
${\rm LOX} = \log(W_{r\lambda}/ 1m{\rm \AA})$, where $W_{r\lambda}$ is the rest equivalent
width, for weak lines. Hence, this quantity can
be used to rank metal lines in terms of strength, and it can be compared to detection limits 
in spectra of a given quality. The model-dependent ionization corrections are contained in the
last term, the ratio of ionization fractions of $Z^i$ and $H^0$ within the absorber. 

\begin{figure}[tb]
\centerline{\vbox{
\psfig{figure=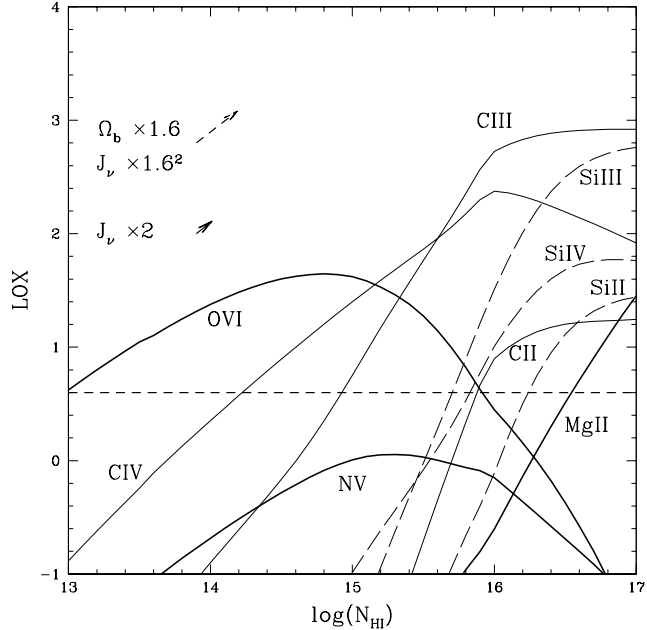,height=9cm}
}}
\caption[]{LOX as a function of HI column density for 9 selected absorption lines from 
carbon (CIV(1548), CIII(977), CII(1335), thin solid lines), silicon (SiIV(1394), SiIII(1206),
SiII(1260), dashed lines), and OVI(1032), NV(1243), and MgII(2796) (bold solid lines). The
solid arrow indicates the direction and magnitude of changes to the curves if $J_{\nu}$
is doubled, keeping everything else constant, while the dashed arrow indicates the effect
of changing the baryonic density while adjusting $J_{\nu}$ to keep the mean
flux decrement of the spectrum constant (see \cite{hel97b}). The horizontal dashed line
indicates LOX=0.6, roughly corresponding to the detection limit in the best available spectra.}
\end{figure}
As a specific application of the LOX technique we use the standard CDM, $\sigma _{8h^{-1}}=0.7,
h=0.5$, $\Omega _b = 0.05$ simulation described in \cite{kat96} and \cite{dav96}. We assume a 
Haardt \& Madau background radiation spectrum of slope -1.8 at the high-energy end \cite{haa97} 
normalized to match the observed mean flux decrement $D_A$  \cite{pre93}, and we focus on
redshifts $z \sim 3$. The simulations exhibit rather tight correlations between $N_{\rm HI}$ 
of an absorbing region and the typical temperature and total gas density within the region, 
so the ionization correction term is essentially a function only of $N_{\rm HI}$, and we 
calculate this term using the photoionization code CLOUDY 90 \cite{fer96}. 

The LOX has been evaluated for 199 metal lines from the database contained in the Voigt
profile fitting software VPFIT \cite{car87}. Figure 1 shows the results as a function on $N_{\rm HI}$
for nine of the strongest lines, assuming a carbon abundance of [C/H]=-2.5 and a relative abundance pattern similar to that
observed in population II stars. A more exhaustive list of lines is presented in \cite{hel97b}, 
which also discusses the dependence of the LOX on $z$, $\Omega _b$, and the normalization of the
radiation field $J_{\nu}$. This dependence is weak, and the following predictions are believed to hold
rather generally in cosmological models of the \lya forest:

\begin{itemize}
\item{The CIV(1548 \AA, 1550 \AA) doublet is the strongest line with rest wavelength $\lambda _r >
1216 {\rm \AA}$ regardless of the HI column density of the absorber.}
\item{OVI(1032 \AA, 1038 \AA) is the only detectable line in low column density 
($\log{N_{\rm HI}} \simlt 14.5$) absorbers, and hence the best probe of metallicity in the
low density IGM.}
\item{The CIII(977 \AA) and SiIII(1206 \AA) lines are the potentially strongest metal lines, but
they are only expected to be seen in the relatively uncommon absorbers with $N_{\rm HI} \simgt 10^{16}
{\rm cm}^{-2}$.}
\end{itemize}
 
The first prediction is easily verifiable from available data and is found to hold. The other
predictions, featuring lines that are embedded in the \lya forest, remain to be verified.
If few or no OVI lines are found at $z \sim 3$ in real spectra then the
metallicity of the low-density regions of the IGM may actually be less than -2.5, or alternatively
there may be a problem with this picture of the \lya forest. In absorbers with higher
values of $N_{\rm HI}$, the higher number of observable metal lines makes a more thorough 
comparison between model and observations possible. We are currently initiating a detailed systematic 
study of the distributions and relative strenghts of metal lines associated with such systems. 

\acknowledgements{We have benefitted from discussions with Chris Churchill.
UH acknowledges support by a postdoctoral research grant from the
Danish Natural Science Research Council. This work was supported by the NSF under grant
ASC93-18185 and the Presidential Faculty Fellows Program
and by NASA under grants NAG5-3111, NAG5-3525, and NAG5-3922.
Computing support was provided by the San Diego Supercomputer Center. 
 }


%
\vfill
\end{document}